\documentclass[final,5p,times,twocolumn,numbers,sort&compress]{elsarticle}

\usepackage{pifont} 
\usepackage{amssymb}
\usepackage{lipsum}
\usepackage{lineno}

\usepackage[dvipsnames]{xcolor}
\definecolor{lightteal}{HTML}{20B2AA}

\definecolor{darkteal}{HTML}{008B8B}
\usepackage[colorlinks=true,
            linkcolor=darkteal,
            citecolor=darkteal,
            urlcolor=darkteal,
            pdfborder={0 0 0}]{hyperref}
\usepackage{hypernat}

\newcommand{\xmark}{\ding{55}}

\journal{Physics Letters B}
\biboptions{square,comma,sort&compress}
\begin{document}
%\linenumbers
\hypersetup{
  colorlinks=true,
  allcolors=darkteal
}
\begin{frontmatter}
\title{Implications of a Weakening $N=126$ Shell Closure Away from Stability for $r$-Process Astrophysical Conditions}

\author[first,second,third]{Mengke Li}
\ead{mengkel@berkeley.edu}
\cortext[cor1]{Corresponding author:}
\author[first,fourth]{Gail C. McLaughlin}
\author[first,third]{Rebecca Surman}

\affiliation[first]{organization={Network for Neutrinos, Nuclear Astrophysics, and Symmetries (N3AS), University of California, Berkeley}, 
            city={Berkeley},
            postcode={94720}, 
            state={CA},
            country={USA}}

\affiliation[second]{organization={Department of Physics, University of California, Berkeley},
            city={Berkeley},
            postcode={94720}, 
            state={CA},
            country={USA}}

\affiliation[third]{organization={Department of Physics and Astronomy, University of Notre Dame},%Department and Organization
            city={Notre Dame},
            postcode={46656}, 
            state={IN},
            country={USA}}
            
\affiliation[fourth]{organization={Department of Physics, North Carolina State University},%Department and Organization
            city={Raleigh},
            postcode={27695}, 
            state={NC},
            country={USA}}

\begin{abstract}
The formation of the third $r$-process abundance peak near $A \sim 195$ is highly sensitive to both nuclear structure far from stability and the astrophysical conditions that produce the heaviest elements. In particular, the $N=126$ shell closure plays a crucial role in shaping this peak. Experimental data hints that the shell weakens as proton number departs from $Z=82$, a trend largely missed by global mass models. 
To investigate its impact on $r$-process nucleosynthesis, we employ both standard global models with strong closures and modified Duflo–Zuker (DZ) models that reproduce the weakening, combined with three sets of $\beta^-$-decay rates. Strong shell closures generate sharply peaked abundances, whereas weakened closures consistent with the experimental trend produce broader, flatter patterns. Accurately reproducing the solar third peak under weakened shell strength requires highly neutron-rich conditions ($Y_e \leq 0.175$) and slower decay rates.
These results demonstrate that a weakening $N=126$ shell closure away from stability imposes significant constraints on the astrophysical environments of the $r$-process and underscores the need for precise mass measurements and improved characterization of $\beta^-$-decay properties in this region.
\end{abstract}

\begin{keyword}
r-process nucleosynthesis \sep N = 126 Closed shells \sep Nuclear Masses \sep Neutron star mergers
\end{keyword}

\end{frontmatter}

\section{Introduction}
\label{introduction}

The origin of the elements remains one of the central questions in modern astrophysics. While light elements up to iron are primarily synthesized through stellar fusion processes, the heaviest elements in the periodic table are thought to be produced in environments with extreme neutron fluxes. The rapid neutron-capture process ($r$-process) has long been recognized as the dominant mechanism responsible for creating roughly half of the elements heavier than iron in the Universe \citep{BBF_1957}. Despite its importance, a comprehensive understanding of the $r$-process remains challenging, as it requires the interplay of astrophysical modeling, astronomical observations, and detailed knowledge of the nuclear properties of the synthesized nuclei. 

On the astrophysical side, significant progress has been made in constraining the potential $r$-process sites. Binary neutron star mergers have been confirmed through multi-messenger observations \citep{GW_2017, Pian_2017}. However, observations of metal-poor stars suggest that it remains uncertain whether neutron star mergers occur frequently enough, or early enough in galactic history, to explain all of the observed data \citep{2004_Argast, 2014_Komiya,2014_Matteucci,Kobayashi_2023}. Because of their relatively delayed timescales, additional astrophysical environments have been proposed as potential contributors to $r$-process nucleosynthesis \cite{Cowan_2021}. These include, but are not limited to, neutron star–black hole mergers \cite{Curtis_2023}, magneto-rotational core-collapse supernovae \citep{Nishimura_2015, Mosta_2018}, accretion-disk outflows from collapsars \citep{Siegel_2019}, and recently magnetar giant flares \citep{Patel_2025}.

From the nuclear physics perspective, the challenge arises because the nuclei involved in the $r$-process are extremely neutron-rich and short-lived, making them very difficult, and in many cases impossible, to measure directly in the laboratory. Among the relevant nuclear properties, nuclear masses are the most fundamental, as they are required for nearly all reaction-rate calculations. Over the past decades, impressive progress has been made in reaching nuclei closer to the $r$-process path at radioactive beam facilities worldwide, such as RIBF at RIKEN \citep{RIBF_2012}, CARIBU at Argonne \citep{CARIBU_2008} and TITAN at TRIUMPH \citep{TITAN_2024}. Nevertheless, many regions remain experimentally inaccessible, and nuclear masses in these areas must still be estimated theoretically to enable $r$-process studies.

Several theoretical approaches have been developed to predict nuclear masses in regions inaccessible to experiment. Macroscopic–microscopic models, such as the finite-range droplet model (FRDM \cite{FRDM_2012}), combine a deformed droplet model that captures bulk nuclear properties with shell corrections that reproduce the observed shell closures, while self-consistent methods like Hartree–Fock–Bogoliubov (HFB \cite{HFB_2013}) and energy density functional (EDF \cite{DFT_2003}) theory provide a 
%more 
microscopic framework. The Duflo–Zuker (DZ \cite{DZ_1995}) model, rooted in shell-model monopole interactions, offers a compact phenomenological description of global masses. Ab initio methods are beginning to reach medium-mass nuclei \citep{Ab_2013}, and recent machine-learning techniques provide complementary, data-driven extrapolations \citep{ Jorge_2017, PIML_2022, PIML_R_2024}.

These theoretical efforts are essential because nuclear masses directly determine neutron separation energies, which in turn govern the location of the $r$-process path. One of the most striking manifestations of this connection is the appearance of pronounced abundance peaks in the solar $r$-process distribution near mass numbers $A \sim 80, 130$, and $195$. These peaks provide direct evidence of the influence of nuclear structure, as they are associated with closed neutron shells at $N = 50$, 82, and 126. At these so-called “waiting points” \citep{Waiting_point_1983}, large shell gaps suppress neutron captures, causing material to accumulate and shaping the final abundance distribution. Among them, the shell closure at $N = 126$ is of particular importance, as it governs the formation of the third $r$-process peak near $A \sim 195$. This peak is not only one of the most prominent features in the solar $r$-process abundance pattern, but it also corresponds to the region that includes many of the heaviest stable elements, such as gold and platinum. Despite its critical role, the nuclear structure in this region remains poorly constrained experimentally, since most of the relevant isotopes lie far from stability and are not yet accessible with current facilities. Consequently, theoretical nuclear mass models are indispensable for guiding $r$-process simulations.

A convenient measure of the $N = 126$ shell closure is its strength, which can be quantified using the two-neutron shell gap \citep{Brown_2022}, defined as
\begin{equation}
D_{2n}(Z, N) = S_{2n}(Z, N) - S_{2n}(Z, N+2),  
\end{equation}
where $S_{2n}$ is the two-neutron separation energy. A large $D_{2n}$ reflects a robust shell gap, while a reduction signals a weakening of the shell closure. 
In the $N = 126$ region, significant experimental progress has been made in recent decades, with precision mass measurements from facilities such as ISOLTRAP at CERN and storage-ring experiments at GSI providing valuable data \citep{GSI_2006}. 
Compiled data from the Atomic Mass Evaluation (AME) \citep{AME2020} indicates that the $N = 126$ shell strength decreases both above and below the doubly magic proton number $Z = 82$. Despite these advances, the data remain sparse, particularly for nuclei with $Z < 82$ where only two data points are available \citep{Hg_208, Pb_208}, and the systematic behavior of the shell gap in this critical region is not yet fully constrained.

To overcome the lack of experimental data in this region, it is essential to compare predictions from different theoretical mass models. Global models such as FRDM, HFB, and DZ provide varying estimates for the evolution of the $N = 126$ shell strength far from stability, allowing us to assess the impact of nuclear structure on $r$-process abundances.
While FRDM and HFB generally predict a strong and relatively constant shell closure, the DZ model shows a decreasing trend qualitatively similar to experimental data, albeit with a less pronounced evolution. This diversity of predictions raises two key questions: (1) How do different $N = 126$ shell strengths influence the formation of the third $r$-process peak under various astrophysical conditions? (2) If the true $N = 126$ shell strength follows the decreasing trend suggested by AME data, what can it reveal about the astrophysical sites and conditions responsible for producing the third $r$-process peak?

In this work, we first examine the predictions of global mass models (FRDM, HFB, and DZ) for the $N=126$ shell closure. We then modify the DZ model to create two distinct shell-strength scenarios: a weakened closure, consistent with the decreasing trend indicated by AME data, and a stronger closure, comparable to FRDM and HFB predictions. Using these mass tables, we study their impact on $r$-process simulations, with a particular focus on the formation of the third abundance peak. Details of the model modifications and simulation setup are provided in the Methods section, while the resulting abundance patterns and their astrophysical implications are presented in the Results section. This approach allows us to directly assess how nuclear structure far from stability influences the conditions necessary for producing the heaviest elements.

\section{Methods}
To explore the evolution of the $N = 126$ shell closure, we began with the DZ mass model, which already exhibits a decreasing trend in shell strength away from $Z = 82$. The DZ model includes 33 parameters, of which 31 are nonzero. The shell strength arises from the interplay of multiple coupled terms rather than a single coefficient. These terms collectively shape the mean field, influence shell gaps, and determine how magic numbers evolve as nuclei move away from stability. To identify those with the largest impact on the $N=126$ shell closure, we varied each coefficient individually by factors of 1.2 and 0.8 and examined the resulting changes in the two-neutron separation energy, $S_{2n}$, a direct indicator of shell strength. Among them, the $17^{th}$ coefficient—associated with the $SQ^-$ term, a monopole–quadrupole component that encodes how extruder and intruder orbitals rearrange across major shells—proved particularly influential. Increasing this term produced a weaker closure consistent with the AME trend, while decreasing it yielded a stronger shell similar to FRDM and HFB predictions.

Since this study focuses on the third $r$-process peak, we aimed to minimize deviations from the baseline DZ masses outside the region of interest. To achieve this, we combined the default DZ mass model with the modified version at $N = 110$. This choice ensures continuity, as the modified parameter does not affect the predicted $S_{2n}$ values at $N = 110$, thereby eliminating potential edge effects when employing the full data set.

%The 
We employ the combined mass table 
%was employed 
in $r$-process simulations as follows. One-neutron separation energies, $S_{1n}$, are updated accordingly, as they exponentially affect the photon dissociation rates. To account for theoretical uncertainties in $\beta^-$-decay rates, we examine three widely used models. Möller et al. \citep{MLR_2019} (hereafter MLR) employ a finite-range droplet model (FRDM) combined with the quasiparticle random phase approximation (QRPA) to calculate $\beta^-$-strength functions for neutron-rich nuclei. Ney et al. \citep{Ney_2020} (hereafter Ney) use the finite amplitude method with Skyrme density functionals to compute $\beta^-$-decay half-lives across the neutron-rich region. Finally, Marketin et al. \citep{MKT_2016} (hereafter MKT) apply a covariant density functional theory framework to evaluate the $\beta^-$-decay rates of r-process nuclei.
Together, these models provide complementary perspectives to incorporate the impact of $\beta^-$-decay rate uncertainties on nucleosynthesis predictions, while all other reaction rates are taken from the default PRISM reaction network \citep{Mumpower_2018}.
 
For this work, we focus on so-called `hot’ astrophysical conditions \cite{Mumpower_2016}. In a classic hot $r$-process, an extended ($n,\gamma$)-($\gamma,n$) equilibrium phase is followed by a freeze out phase where neutron capture, beta decay, and photodissociation all compete to finalize the $r$-process abundance pattern. This is in contrast to a `cold’ r-process, where the temperature drops quickly, ($n,\gamma$)-($\gamma,n$) equilibrium fails early, and beta decay and neutron capture alone shape the final abundances. Since photodissociation rates depend exponentially on the neutron separation energies, nuclear masses have the greatest leverage on a hot $r$-process, so for this first study we confine our discussion to these types of astrophysical conditions. We adopt the same set of hot trajectories as in \citep{Traver_2020}, consisting of representative tracers from both dynamical and wind ejecta extracted from simulated disk outflows \citep{just_2015}. The compositions are then parameterized to span electron fractions $Y_e$ from 0.05 to 0.35 and entropies per baryon $s/k = 20$ and 40, sampling a reasonable range of astrophysical conditions. With appropriate weighting, this selection can reproduce the solar $r$-process abundance pattern in \citep{Traver_2020}. We adopt this weighting for the present work. We look forward to future work where we propagate the nuclear structure effects of varying neutron shell closure strengths to neutron capture cross sections and beta decay rates in order to include cold $r$-process simulations in our analysis.

\section{Results}
With the framework established, we now present our results, beginning with the $N = 126$ shell strength from $D_{2n}$, followed by the impact of mass models and $\beta^-$-decay prescriptions on the $r$-process abundances.

\subsection{$N = 126$ closed shell strength}
\begin{figure}[h!]
    \centering
    \includegraphics[width=0.95\linewidth]{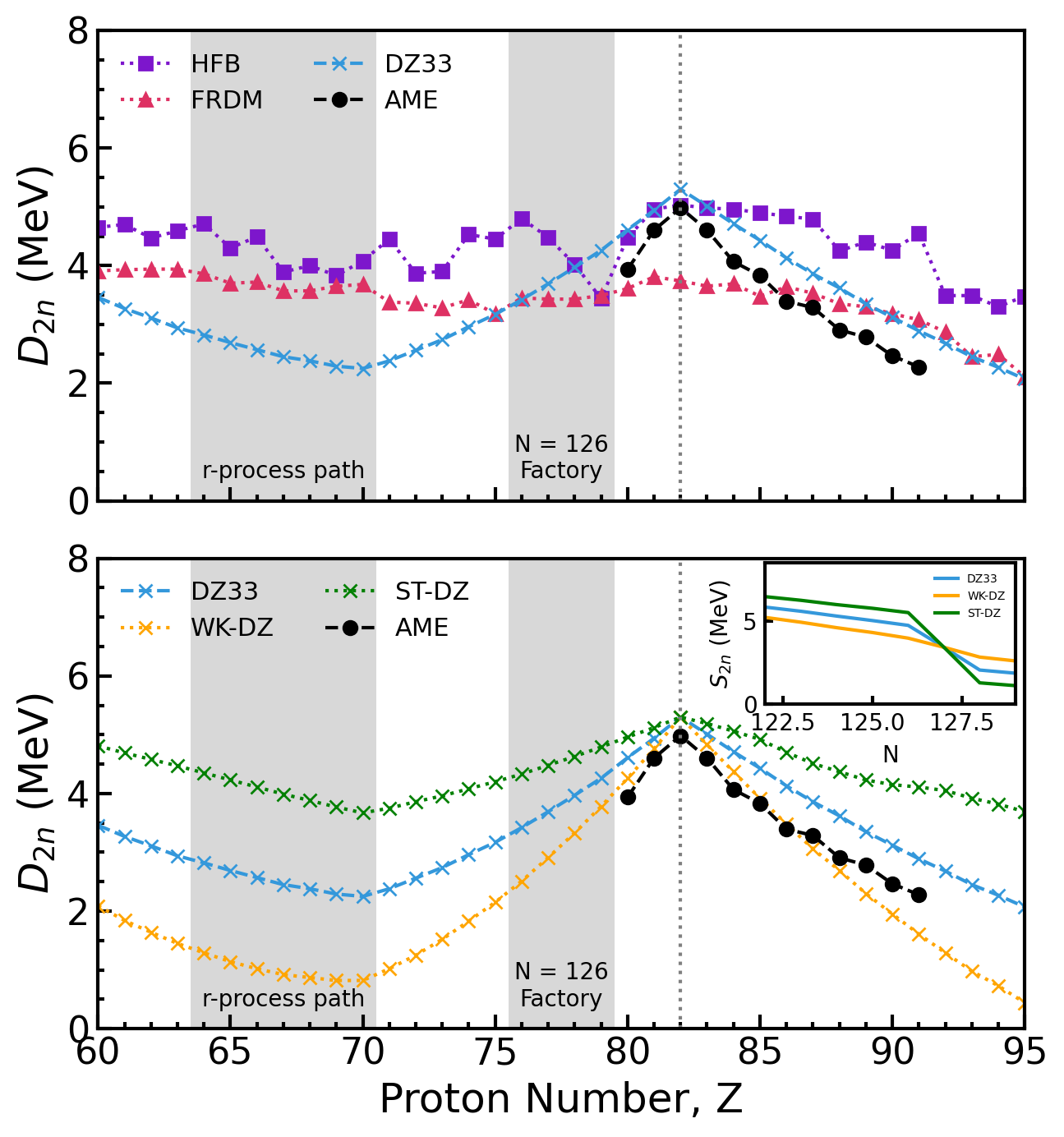}
    \caption{Strength of the $N = 126$ shell closure from different mass models. The upper panel shows the two-neutron shell gap $D_{2n}$ predicted by HFB (purple), FRDM (pink), DZ (blue), and experimental AME2020 data (black). The lower panel compares the default DZ with its modified variants: WK-DZ (weaker shell closure) and ST-DZ (stronger shell closure). The inset highlights the $S_{2n}$ values for the $Z = 60$ isotopic chain, illustrating the contrast in shell strength across $N = 126$. The shaded gray regions mark the $r$-process path range and, separately, the nuclei proposed for measurement at the $N = 126$ Factory.}
    \label{fig:d_2n}
\end{figure}

To illustrate the strength of the $N = 126$ closed shell, we show the two-neutron shell gap values from both AME data and theoretical predictions in Figure~\ref{fig:d_2n}. 
The upper panel compares AME data (black dots) with three theoretical models. The experimental points suggest a weakening of the shell closure as the proton number moves away from $Z = 82$. On the larger-$Z$ side ($Z > 82$), where more data are available, the decreasing trend is especially evident; on the lower-$Z$ side, the scarcity of data limits firm conclusions. This motivates a closer look at theoretical predictions to assess how well they capture the observed behavior.

Among the theoretical models, FRDM (pink) predicts nearly uniform shell strength with only slight variation across $Z$, maintaining a stronger closure relative to the trend suggested by AME data for smaller proton numbers along the r-process path (shaded in gray). HFB (purple) follows the AME trend from $Z=80$-82, but similarly yields consistently strong shell closures otherwise, with $D_{2N}$ values ranging from 4 to 5 MeV. 
In contrast, the DZ model (blue) does capture the decreasing trend as $Z$ departs from $82$, though its slope is more gentle. The modified DZ with weaker shell closure (orange) in the lower panel agrees well with the AME2020 data on the higher-$Z$ side and follows the experimental trend on the lower-$Z$ side. By comparison, the modified DZ with stronger closure (green) predicts a shell strength as strong as that of FRDM and HFB.

To illustrate these differences between the default DZ and the modified versions, the inset at the upper right corner in the bottom panel of Figure~\ref{fig:d_2n} focuses on the $S_{2n}$ values for the $Z = 60$ isotopic chain. 
This close-up highlights the contrast between the default DZ model and its modified variants: the weaker-shell case (orange) shows only a mild drop across the closure, while the stronger-shell case (green) exhibits a sharp discontinuity.

\subsection{r-process abundances using different mass models}

\begin{figure}[h!]
    \centering
    \includegraphics[width=0.95
    \linewidth]{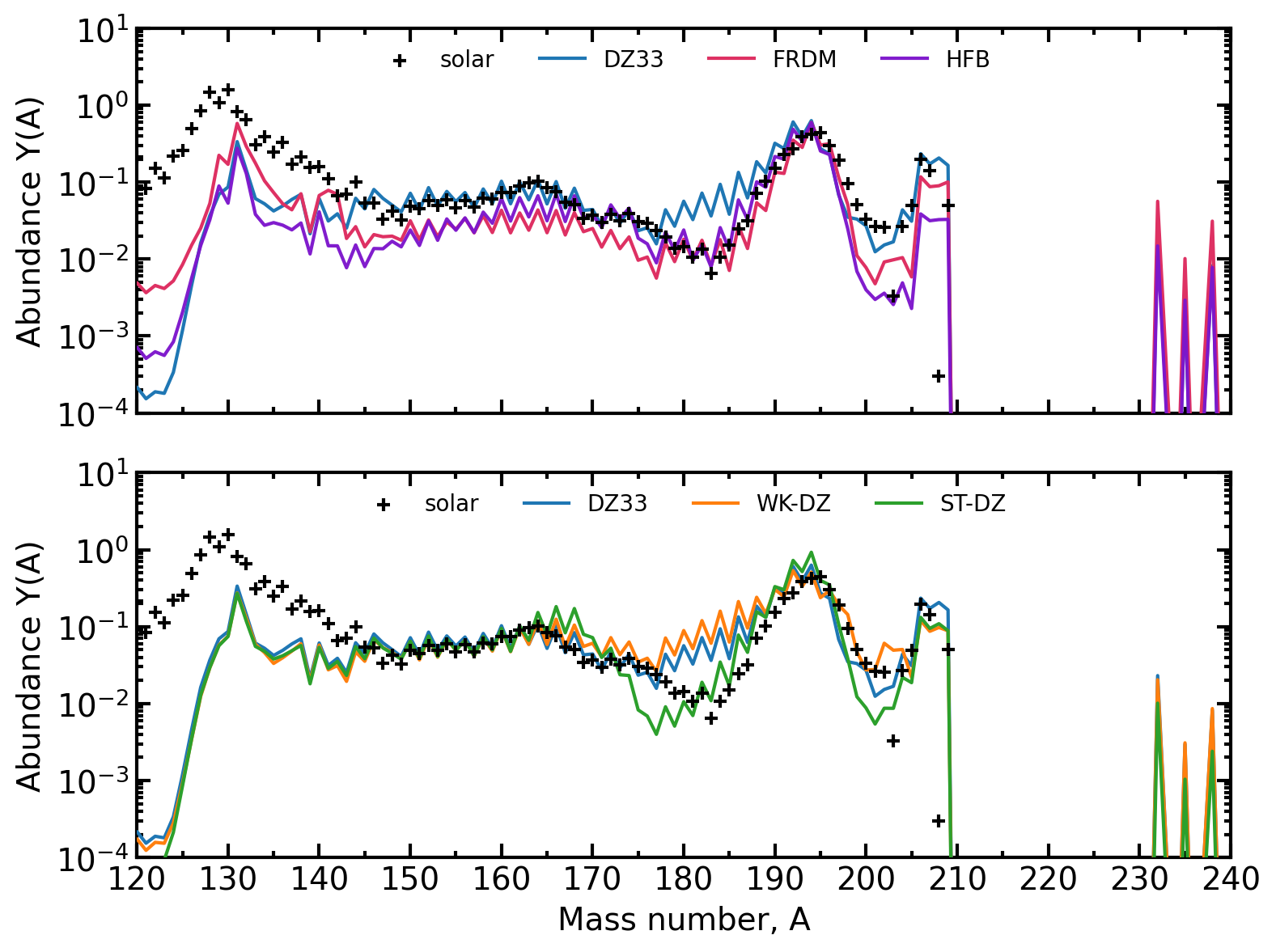}
    \caption{Simulated $r$-process abundance patterns for a single trajectory, with $Y_e = 0.2$ and entropy per baryon $s/k=40$, using different nuclear mass models. The top panel shows the results with various global mass models, while the bottom panel compares results using the default DZ model with its modified versions. Solar $r$-process residuals from \citep{Arnould_2007} are shown as points.}
    \label{fig:ya_diff_masses}
\end{figure}

With different predicted strengths of the $N = 126$ shell closure, the resulting $r$-process abundance patterns are shown in Figure \ref{fig:ya_diff_masses} compared to the solar pattern. The upper panel compares simulations based on three global theoretical mass models. When the shell closure is strong, as predicted by FRDM and HFB, the third $r$-process peak is reproduced well. In contrast, simulations using the DZ model, which exhibits a weakened shell closure consistent with the decreasing trend as seen in the AME2020 data, yield a broader third peak and an overproduction on its lower-mass side. Notably, the default DZ model lies between the extremes: it is neither as weak as the AME trend nor as strong as FRDM and HFB.

To further assess the impact of shell strength on the formation of the third $r$-process peak, the lower panel shows results obtained with the modified DZ mass models. When the shell gap is artificially strengthened to match the magnitude predicted by FRDM and HFB, the simulation reproduces the solar abundance pattern around the third peak with high fidelity (green line). Conversely, when the shell gap is weakened further, in line with the AME2020 trend, the third peak becomes even broader than in the default DZ case. Although this scenario produces an abundance pattern that deviates from the solar pattern, it may represent the true underlying nuclear physics. 
This further underscores the strong sensitivity of the final abundances to the evolution of the $N = 126$ shell closure, highlighting the need for precise mass measurements in this region.

To investigate how the strength of the $N = 126$ shell closure affects the formation of the third $r$-process peak, we first examine the abundance evolution along a single trajectory. In hot astrophysical conditions, the $r$-process path---defined as the most abundant isotopes in each isotopic chain and governed by the Saha equation---follows lines of nearly constant neutron separation energy ($S_n$) on the nuclear chart. As equilibrium persists until the supply of free neutrons is exhausted, the final abundance pattern is largely determined at the end of the $(n,\gamma) \leftrightarrow (\gamma,n)$ equilibrium phase.

\begin{figure}[h!]
    \centering
    \includegraphics[width=0.9\linewidth]{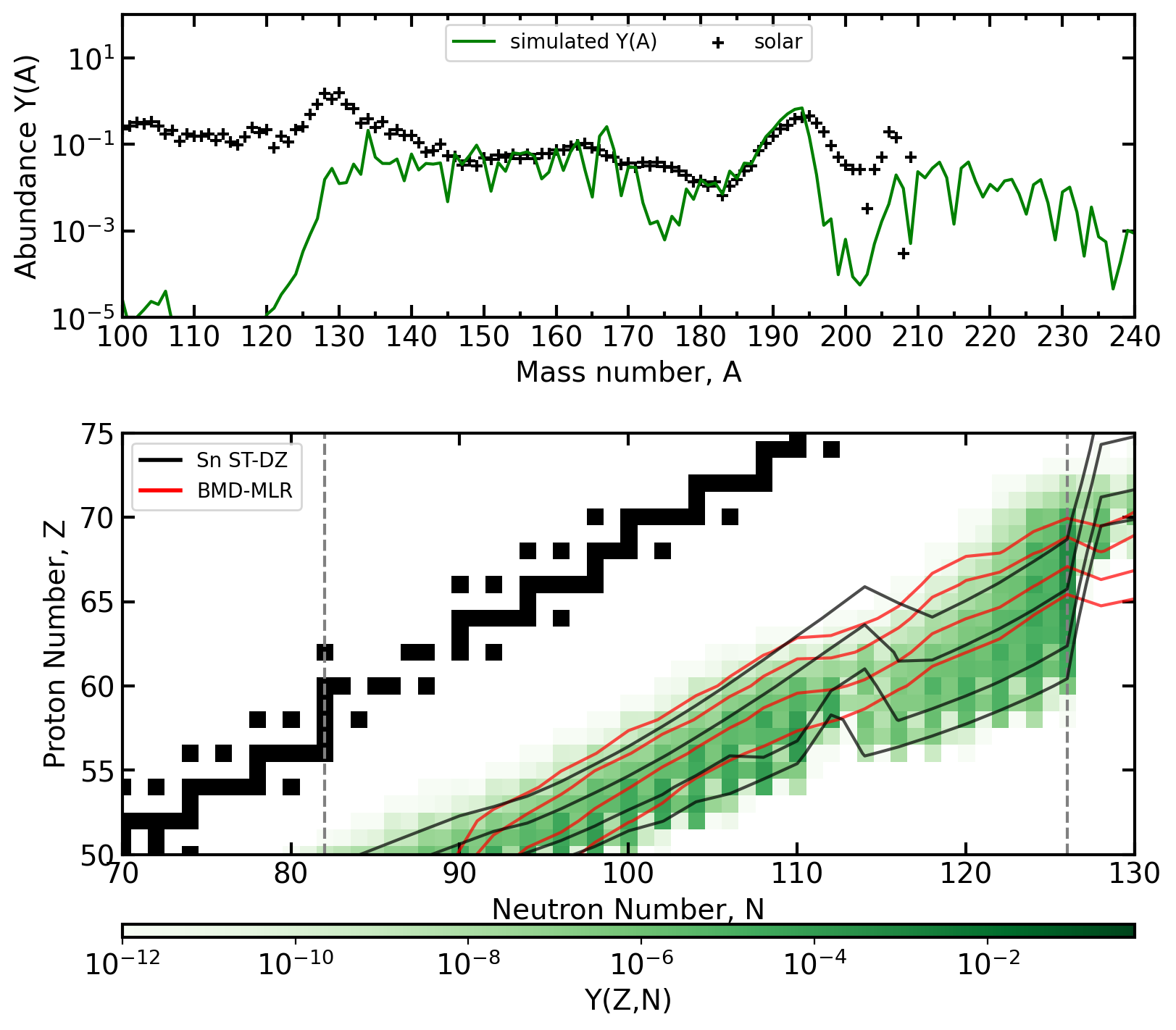}
    \caption{Top: abundance pattern at freeze-out from ($n,\gamma$)–($\gamma,n$) equilibrium with a stronger shell closure, compared to solar $r$-process residuals (black stars). Bottom: isotopic abundances on the nuclear chart, with color indicating relative abundance. Stable nuclei are shown as black squares; black and red contours mark neutron-separation energies and $\beta^-$-decay rates, respectively, and dashed vertical lines indicate neutron closed shells.}
    \label{fig:2p_str}
\end{figure}

\begin{figure}[h!]
    \centering
    \includegraphics[width=0.9
    \linewidth]{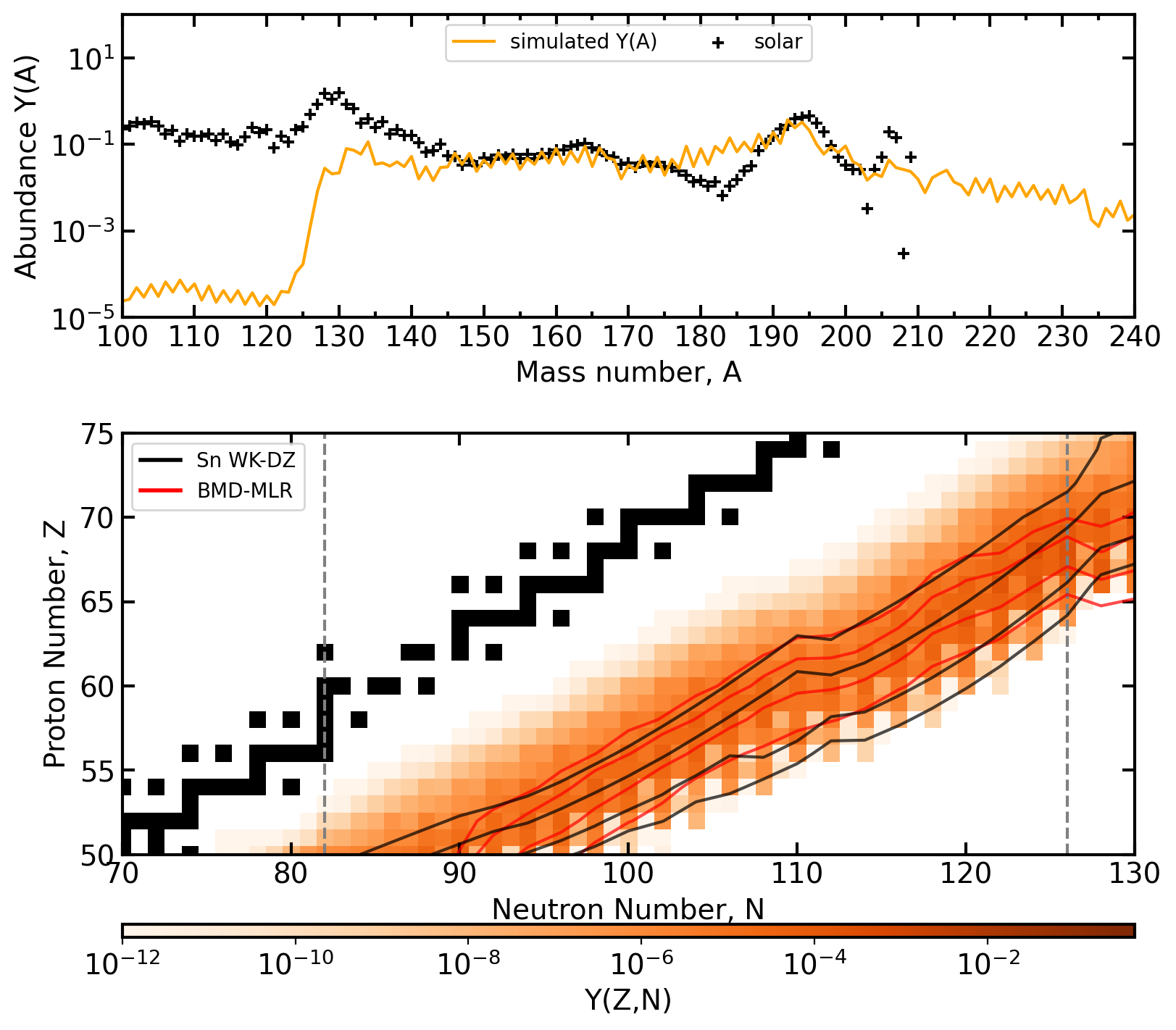}
    \caption{Same as Fig. \ref{fig:2p_str}, but with a weaker shell closure applied in the simulation.}
    \label{fig:2p_wk}
\end{figure}

To visualize this process, Figures~\ref{fig:2p_str} and \ref{fig:2p_wk} present snapshots of the abundance distribution at the end of equilibrium. Figure~\ref{fig:2p_str} shows the strong shell closure case. The upper panel displays abundances as a function of mass number $A$, while the lower panel presents the $Z$–$N$ distribution, with color indicating abundances. The sharp rise of the $S_{n}$ contour (black line) at $N=126$ leads to a clear accumulation of material in the closed shell, which maps to the third $r$-process peak in the upper panel. 
After equilibrium ends, the abundance pattern changes slightly by late time neutron capturing, yielding the final distribution shown in Figure~\ref{fig:ya_diff_masses}.

In contrast, for the case where the shell strength is weakened following the AME trend (Figure~\ref{fig:2p_wk}), the $S_{n}$ contours (black line) increase smoothly, with no pronounced step at $N=126$. In the stronger-shell case, the $r$-process path is sharply pinned at the closed shell, with most material accumulating there. By comparison, the weaker-shell case allows nuclei along the same $Z$ range to span a wider range of neutron numbers before reaching $N = 126$, distributing the abundances across several isotopes. This spreading increases the abundances around $A \sim 185$ and broadens the low-mass side of the third peak, reflecting the more gradual bottleneck at the shell closure. Although a mild shell effect at $N=126$ still causes some accumulation near $A \sim 195$, the resulting peak is noticeably less sharply defined than the solar data.

\begin{figure}[h!]
    \centering
    \includegraphics[width=\linewidth]{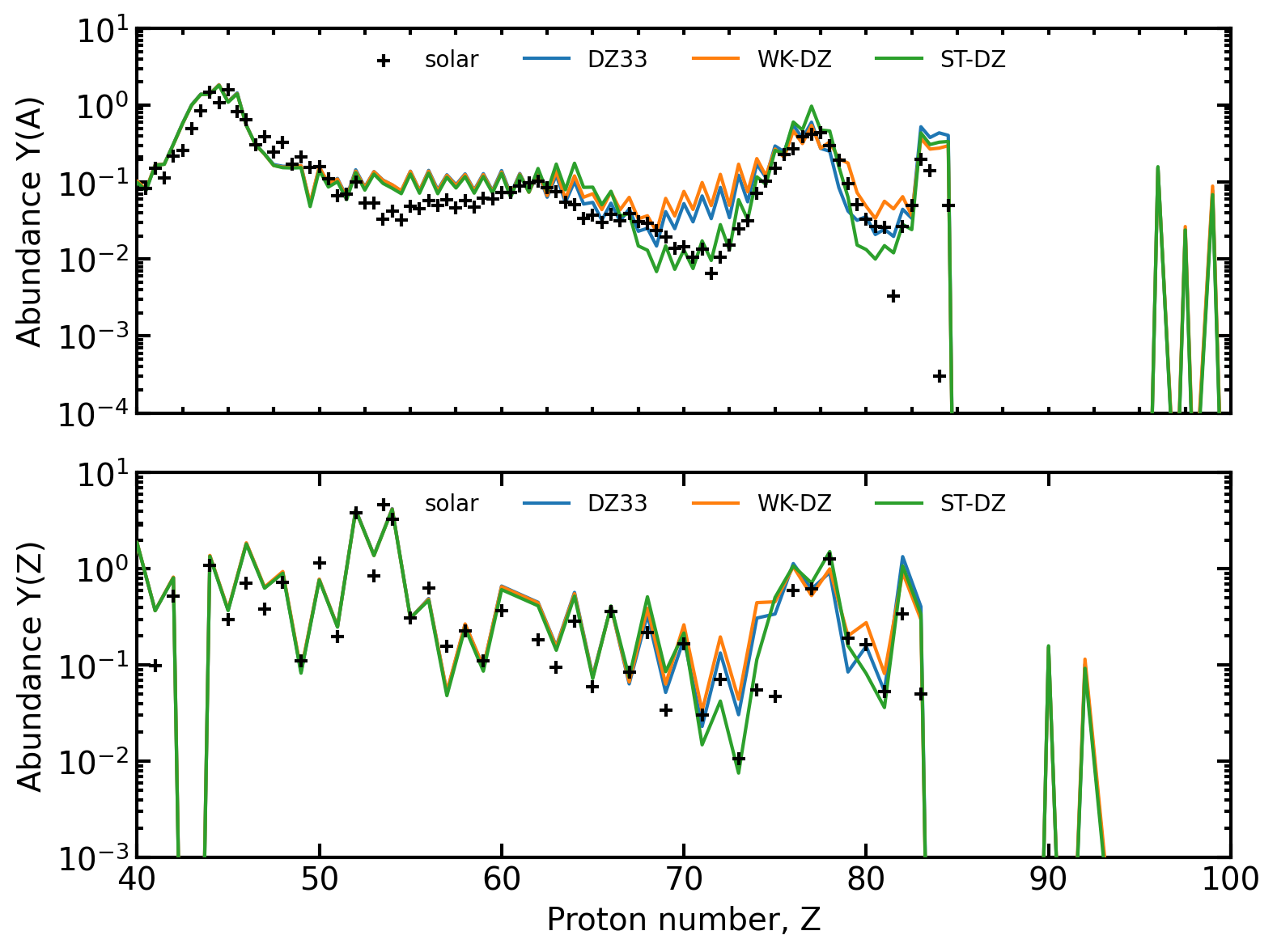}
    \caption{Simulated $r$-process abundance patterns using the default (blue) and modified DZ mass models with weaker (orange) and stronger (green) shell closures. Results are normalized across a set of trajectories to account for varying astrophysical conditions. The upper panel shows isobaric abundances, while the lower panel shows elemental abundances; crosses indicate solar data.}
    \label{fig:ts_moller}
\end{figure}

While the single-trajectory analysis highlights the role of shell strength, it remains unclear whether this behavior persists in more realistic conditions. In astrophysical environments, a distribution of electron fractions ($Y_e$) is expected, and combining contributions from multiple trajectories is essential to reproduce a well-defined third $r$-process peak. To test this, we performed simulations using both the default and modified DZ mass models across
the set of trajectories spanning $Y_e = 0.05$–$0.35$ 
described in the Methods section above. The resulting normalized abundance patterns are shown in Figure~\ref{fig:ts_moller}, with isotopic abundances in the upper panel and elemental abundances in the lower panel.

The multi-trajectory results reinforce the single-trajectory findings. The modified DZ model with stronger shell closure produces a well-defined third peak consistent with the solar data, while the default DZ model yields a broader peak with excess material on the low-mass side. These discrepancies are amplified when the shell strength is further weakened along the AME trend. 
The abundance patterns shown here are calculated using the same trajectory weighting factors adopted in \citep{Traver_2020}. Under these weightings, weaker shell closures fail to reproduce the sharp third peak, and the elemental abundances (lower panel of Figure~\ref{fig:ts_moller}) reflect significant shifts: Hafnium ($Z=72$), Tantalum ($Z=73$), and Tungsten ($Z=74$) are overproduced, while the relative abundances of Gold ($Z=79$) and Mercury ($Z=80$) shift in opposite directions—low gold and high mercury for weaker closures, versus the reverse for stronger ones. To recover a well-defined peak in the weakened-shell case, the contribution from higher-$Y_e$ trajectories must be dramatically reduced, with their weighting lowered to roughly $1$ percent of the original values.

Overall, these results confirm that the strength of the $N = 126$ shell closure is a decisive factor in shaping the third $r$-process peak. When the shell closure is 
strong, the simulations reproduce the solar peak well.
In contrast, a weaker closure consistent with the AME trend produces a broader peak that does not match the solar abundances as closely, yet it may more accurately reflect the true nuclear structure far from stability.

\subsection{Constraint on astrophysical conditions under weakened shell closures}

If the actual shell strength follows the weaker trend suggested by AME, a key question is under what astrophysical conditions the third $r$-process peak can still form without overproduction on the lower mass side while also well positioned. To address this, we examine individual trajectory simulations for electron fractions $Y_e \le 0.25$, since higher $Y_e$ trajectories are insufficiently neutron-rich to produce a third peak and primarily contribute to the second peak and rare-earth region.

\begin{figure*}[!ht]
    \centering
    \includegraphics[width=0.95\linewidth]{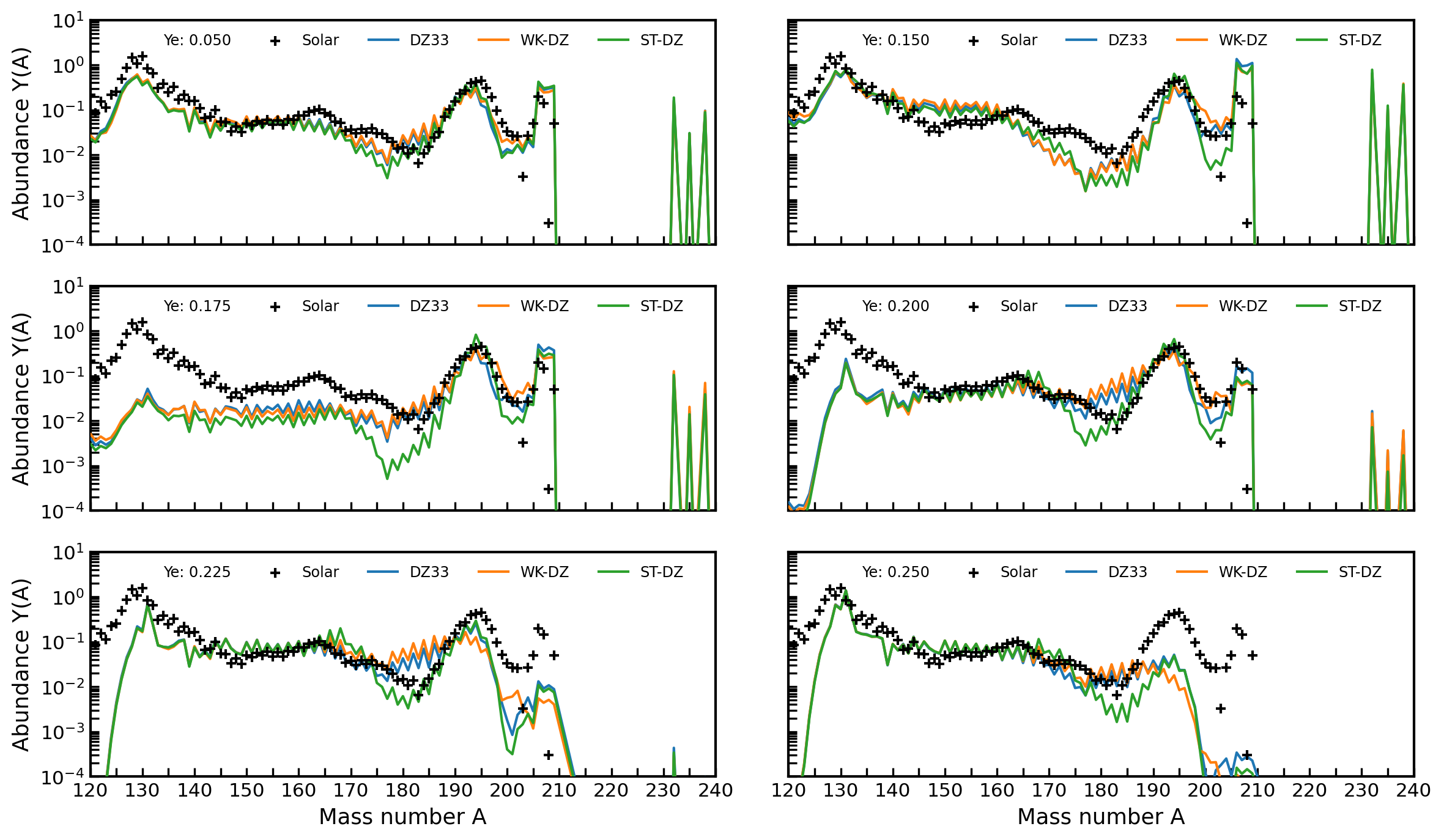}
    \caption{Simulated $r$-process abundance patterns in selected trajectories with varying $Y_e$, using the default (blue) and modified DZ mass models with weaker (orange) and stronger (green) shell closures. Crosses denote solar data.}
\label{fig:diff_ye_moller}
\end{figure*}

Figure~\ref{fig:diff_ye_moller} presents the resulting abundance patterns for representative trajectories. For very neutron-rich conditions ($Y_e \lesssim 0.175$), the third peak remains sharp and closely aligned with the solar pattern. As $Y_e$ increases toward 0.2 and above, weaker shell closure models produce a broader, flattened third peak that shifts slightly to lower mass numbers, deviating from the solar pattern.

This behavior arises from the role of neutron flux. For $Y_e \lesssim 0.175$, the environment is sufficiently neutron-rich for nuclei to capture neutrons beyond the $N=126$ closed shell and form very heavy nuclei that undergo fission, releasing additional neutrons. These fission neutrons are then captured by nuclei just below the third peak, shifting material to higher mass numbers and producing a well-defined structure. In contrast, For $Y_e \gtrsim 0.2$, the neutron supply is limited, so few fissioning nuclei are produced and very small amount or no extra neutrons are released. As a result, the third peak remains at its initial formation position. Under these conditions, a weaker shell closure allows the $r$-process path to populate a broader range of nuclei at lower mass numbers, no further late time neutron capture, resulting in a flatter peak.

Overall, these results indicate that if the weakened shell strength suggested by AME is correct, only moderately to very neutron-rich environments ($Y_e \lesssim 0.175$) can produce a third peak consistent with solar observations, providing a direct constraint on plausible astrophysical conditions.

\subsection{Dependence of the $\beta^-$ decay rates}

The weakened $N=126$ shell strength narrows the range of electron fractions that can produce a well-defined third $r$-process peak. A second critical factor is late-time neutron capture, which depends sensitively on $\beta^-$-decay rates. These rates govern the timing of decays and, consequently, whether nuclei can capture additional neutrons before decaying back to stability. To test the robustness of our conclusions beyond the MLR set, we also examine two widely used alternatives---MKT and Ney---and assess their impact on the resulting abundance patterns.

\begin{figure}[h!]
    \centering \includegraphics[width=\linewidth]{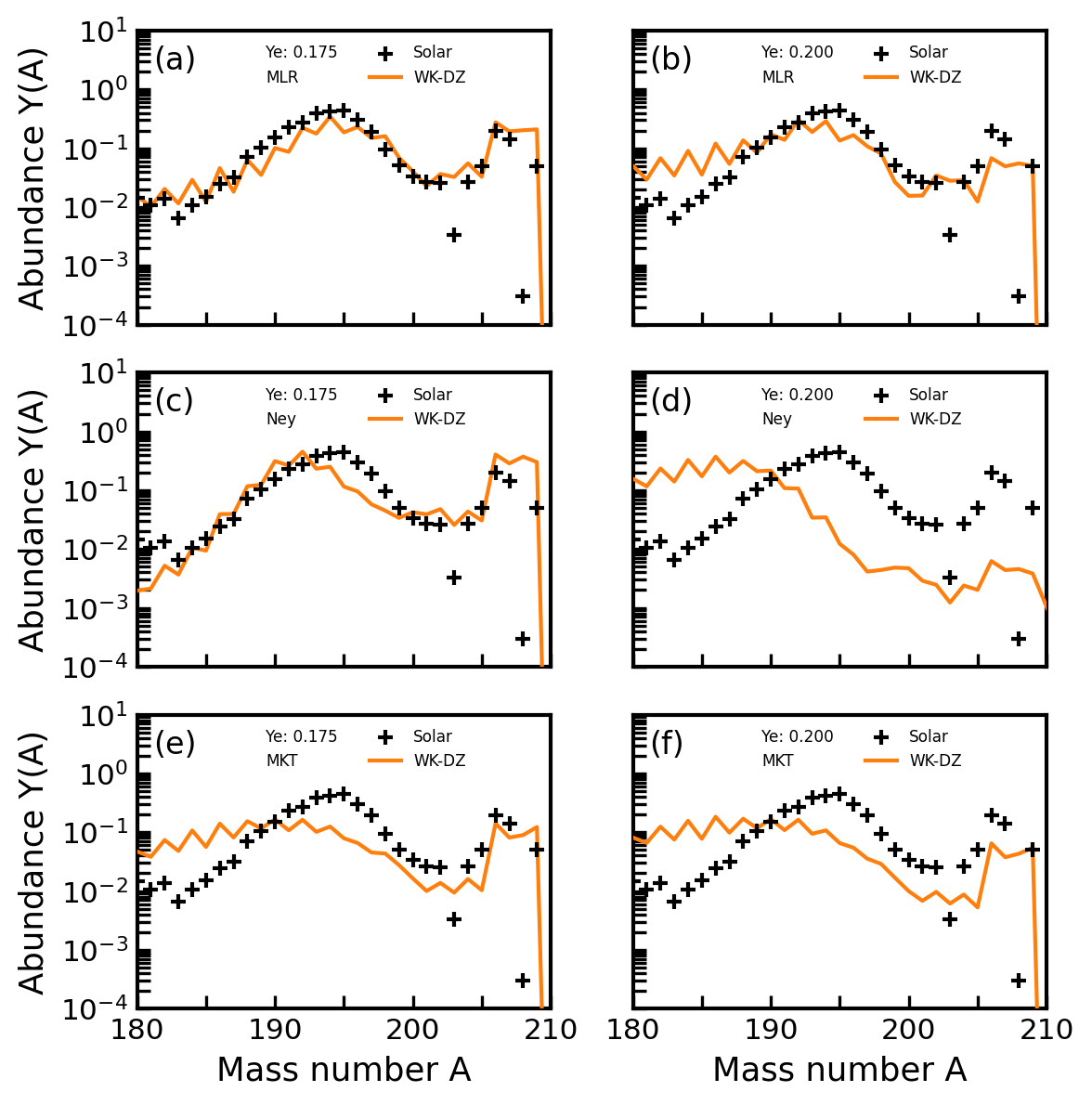}
    \caption{Simulated third $r$-process abundance peaks using the modified DZ mass model with weaker shell closure combined with three $\beta^-$-decay prescriptions: MLR (top), Ney (middle), and MKT (bottom). Results are shown for trajectories with $Y_e = 0.175$ and $0.20$, with solar data indicated by crosses.}
    \label{fig:three_bmd}
\end{figure}

Figure~\ref{fig:three_bmd} compares the simulated third $r$-process peaks obtained with different $\beta^-$-decay prescriptions at two representative $Y_e$ conditions ($0.175$ and $0.20$). The top panel shows results with the MLR set, used here as a reference. The middle panel presents simulations with the Ney set which generally predicts slower decay rates. The main conclusion remains: for $Y_e \leq 0.175$, the combination of slower rates and weaker shell closures produces a well-structured, sharp third peak. At $Y_e \geq 0.20$, however, the peak becomes broader and shifts noticeably to lower mass numbers. In a different manner, the MKT set, which predicts systematically faster decays, fails to reproduce the solar third peak at the correct location for any $Y_e$. This leftward shift is absent in the MLR case, and a detailed study of its origin is left for future work.

To quantitatively assess the agreement between the simulated third $r$-process peak and the solar abundance distribution, we introduce the peak-to-flank ratio, defined as

\begin{equation}
    R = \frac{Y(A_{\rm flank})}{Y(A_{\rm peak})},
\end{equation}

where $Y(A_{\rm peak})$ denotes the abundance at the peak and $Y(A_{\rm flank})$ is the abundance measured 11 mass units to the left of the peak to capture the characteristic extent of the solar peak. The peak locations in panels (a)–(f) of Figure~\ref{fig:three_bmd} occur at $A = 194,\,192,\,192,\,186,\,190,$ and $186$, respectively. We focus on the left flank of the peak to assess its agreement with the solar profile, 
as underproduction on this side can potentially be compensated by other ejecta components, whereas overproduction cannot be mitigated.

\begin{figure}[h!]
    \centering \includegraphics[width=\linewidth]{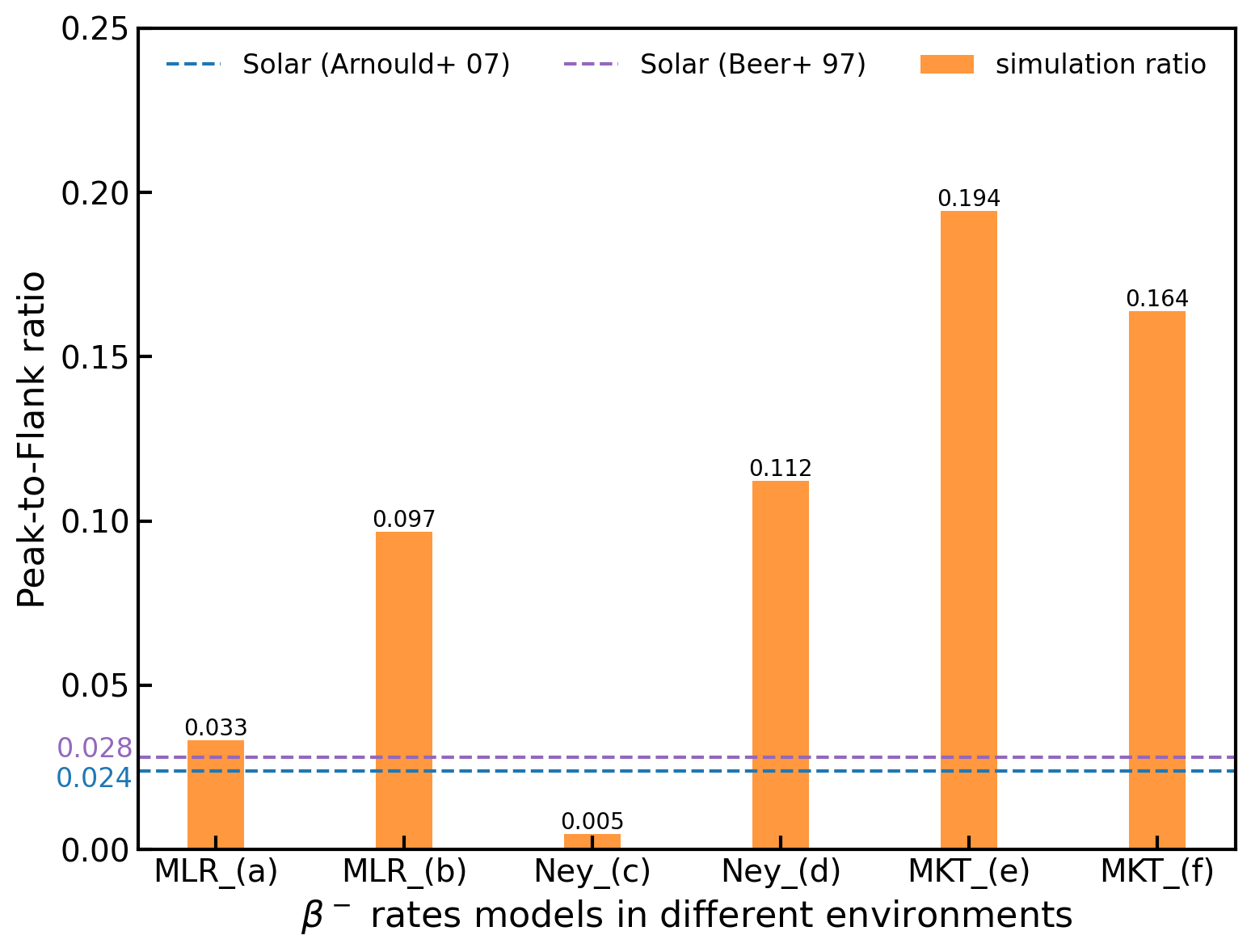}
    \caption{Peak-to-flank ratios ($R = Y(A_{\rm flank})/Y(A_{\rm peak})$) for simulated third $r$-process peaks compared to the solar pattern. Solar reference ratios are 0.023 from \citep{Arnould_2007} (dashed blue) and 0.028 from \citep{Beer_1997} (dashed purple). The x-axis labels indicate the $\beta^-$-decay models and $Y_e$ conditions corresponding to panels (a)–(f) of Fig.~\ref{fig:three_bmd}. Peaks are taken directly from the simulations, with flanks defined 11 mass units below the peak. Ratios near or below the solar values indicate sharper peaks, while larger ratios indicate broader peaks.}
    \label{fig:peak-to-flank}
\end{figure}

The resulting peak-to-flank ratios are shown in Figure~\ref{fig:peak-to-flank}. 
The solar reference values are derived from two sources: Arnould et al. \citep{Arnould_2007} report a ratio of 0.023, while Beer et al. \citep{Beer_1997} give a value of 0.028.
For $Y_e = 0.175$, simulations employing the MLR and Ney $\beta^-$-decay rate models yield ratios comparable to the solar value, with slightly smaller values indicating a sharper peak that remains consistent with observations. In contrast, all other cases produce systematically larger ratios, signifying broader peak structures regardless of their precise location.

\begin{figure*}[!ht]
    \centering
    \includegraphics[width=0.95\linewidth]{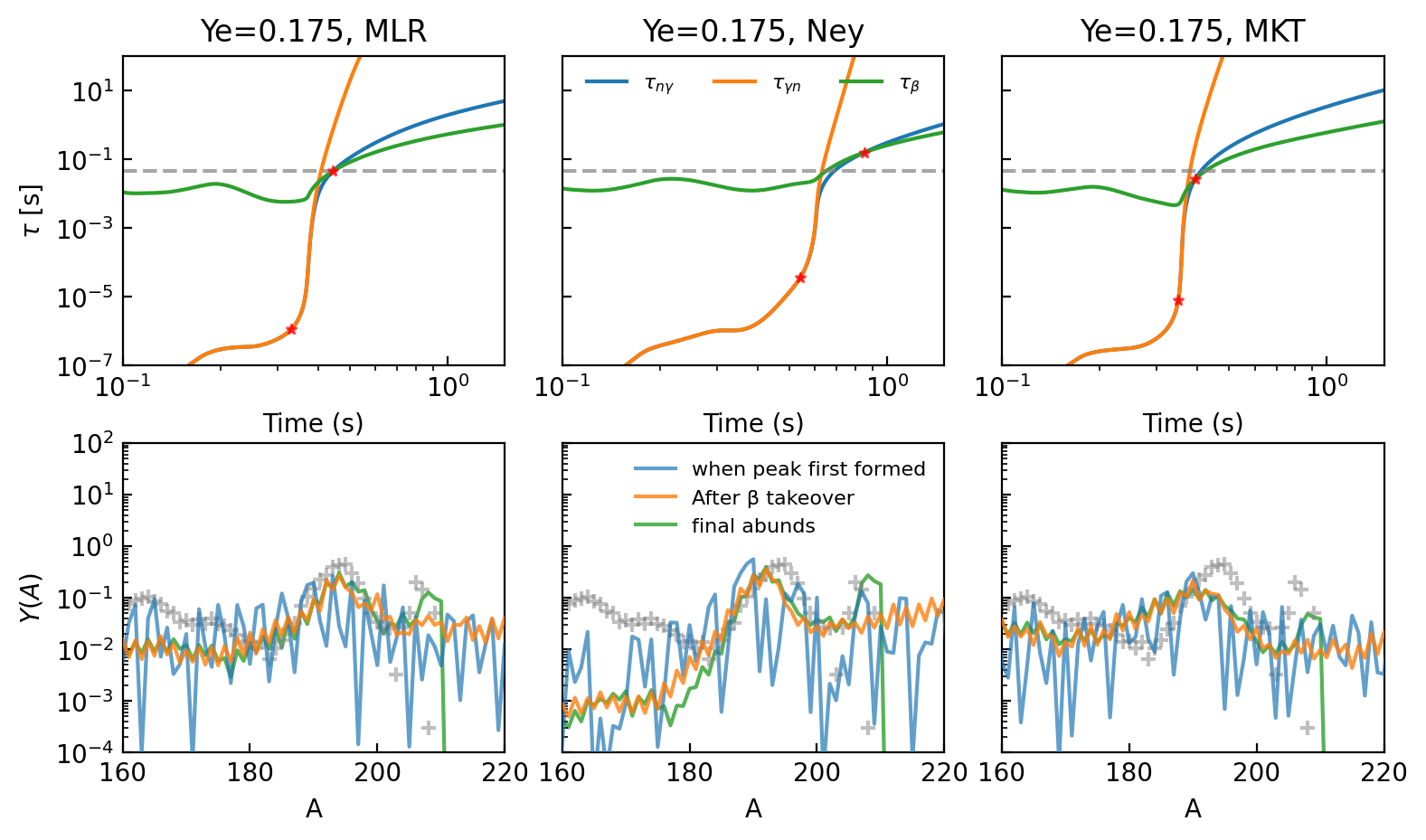}
    \caption{Top: abundance-weighted timescales $\tau$---blue for neutron capture, orange for photodissociation, and green for $\beta^-$ decay---with red stars indicating when the third peak first forms and when $\beta^-$ decay overtakes neutron capture. Bottom: abundance snapshots at these two critical times: blue for initial peak formation, orange for when $\beta^-$ decay dominates. Solar abundances (gray crosses) and the final pattern (green) are shown for reference.}
    \label{fig:late_neutron_capture}
\end{figure*}

To investigate why MLR and Ney rates can reproduce the third $r$-process peak at low $Y_e$, while MKT rates cannot, we examine late-time neutron capture in Figure~\ref{fig:late_neutron_capture} for $Y_e = 0.175$. The upper panel shows the abundance-averaged timescale, with the two red stars marking when the peak first forms and when beta decay begins to dominate. The lower panel presents the corresponding abundance patterns at these two times (blue: first formation, orange: beta-decay dominance). For reference, the solar data and final abundances are also shown.

The interval between the two red stars represents the period during which neutron capture reshapes the third peak. This timescale is longer for the MLR and Ney rates, reflecting their generally slower beta-decay rates, and results in a rightward shift of the peak. In contrast, MKT rates allow little time for late-time neutron capture, so the peak remains largely fixed. These results demonstrate that the ability to capture late-time neutrons, controlled by the beta-decay rates, is critical for forming a well-positioned third $r$-process peak when a weaker shell closure is adopted under neutron-rich conditions.

In summary, our results demonstrate that the formation of a well-defined third $r$-process peak is tightly controlled by both the $N = 126$ shell strength and the choice of beta-decay rates. For the weakened shell closure considered here, following the AME data trend, only sufficiently neutron-rich environments ($Y_e \leq 0.175$) combined with slower beta-decay rates (MLR or Ney) can reproduce a well-structured third peak, while higher $Y_e$ or faster decay rates (MKT) result in a broader, left-shifted peak inconsistent with solar abundances. We show the summary based on Figure \ref{fig:peak-to-flank} in Table \ref{tab:peak_summary}. These findings highlight that reproducing the third $r$-process peak imposes robust constraints on astrophysical conditions, requiring both an accurate description of nuclear physics near $N = 126$ and low electron fractions. 

\begin{table}[ht]
\centering
\begin{tabular}{lccc}
\hline
\textbf{$Y_e$ Range} & \textbf{MLR} & \textbf{Ney} & \textbf{MKT}\\
\hline
$Y_e \leq 0.175$ & \checkmark & \checkmark & \xmark \\
$Y_e \geq 0.200$ & \xmark & \xmark & \xmark \\
\hline
\end{tabular}
\caption{Summary of third $r$-process peak formation under a weakened $N=126$ shell closure case for different $\beta^-$-decay rate models and $Y_e$ conditions.
Check mark (\checkmark) indicates successful reproduction of the third peak; cross (\xmark) indicates failure.}
\label{tab:peak_summary}
\end{table}

\section{Summary and Discussion}
In this work, we have investigated the impact of the $N = 126$ shell strength on the formation of the third $r$-process peak under hot astrophysical conditions. Using the DZ mass model, we explored modifications to produce both weaker and stronger shell closures and assessed their effect on $r$-process nucleosynthesis across a range of electron fractions ($Y_e$) and beta-decay rate prescriptions (MLR, Ney, and MKT).

Our results show that the formation of a well-defined third $r$-process peak is highly sensitive to both nuclear structure and astrophysical conditions. For the weakened $N = 126$ shell closure---similar to the behavior expected from a straightforward extrapolation of the AME data---only sufficiently neutron-rich environments ($Y_e \leq 0.175$) combined with slower beta-decay rates (MLR or Ney) yield a correctly positioned, well-structured third peak. In contrast, higher $Y_e$ or faster rates (MKT) lead to a broader, left-shifted third peak, which is inconsistent with solar abundances. This highlights that if the $N = 126$ shell closure weakens away from stability, it can serve as a powerful constraint on the astrophysical conditions under which the $r$-process operates.

Looking ahead, upcoming experimental programs, such as precision mass measurements at the Argonne $N = 126$ Factory, will provide direct tests of shell strength in this region. In parallel, multi-messenger observations of $r$-process sites, from stellar abundance surveys to kilonova light curves and spectra, will offer complementary astrophysical constraints. Together, these advances will sharpen the connection between nuclear physics and astrophysics, moving us closer to a consistent picture of how the heaviest elements are formed.

\section*{Acknowledgements}
M.L., R.S. and G.C.M. acknowledge support from the Network for Neutrinos, Nuclear Astrophysics and Symmetries (N3AS), through the National Science Foundation Physics Frontier Center award No. PHY-2020275.
This work was partially supported by the US Department of Energy through contract numbers DE-FG0202ER41216 (G.C.M.), DE-FG0295ER40934 (R.S.), and DE-SC00268442 (ENAF - G.C.M., R.S.), and the Office of Defense Nuclear Nonproliferation Research and Development (DNN $R\&D$), National Nuclear Security Administration, U.S. Department of Energy (G.C.M., R.S.) under contract number LA22-ML-DE-FOA-2440. This research was supported in part by the Notre Dame Center for Research Computing.   M.L., G.C.M. and R.S., thank the Institute for Nuclear Theory for its kind hospitality and stimulating research environment, U.S. Department of Energy grant No. DE-FG02- 00ER41132.

\bibliographystyle{unsrt} 
\bibliography{example}
\end{document}